\documentclass[twocolumn,printnumbers,amsmath,amssymb,showpacs,prl]{revtex4}
\usepackage{graphicx}

\begin{document}

\title{Probing the glass transition from structural and vibrational properties of zero-temperature glasses}

\date{\today}

\author{Lijin Wang}
\author{Ning Xu$^*$}

\affiliation{CAS Key Laboratory of Soft Matter Chemistry, Hefei National Laboratory for Physical Sciences at the Microscale and  Department of Physics, University of Science and Technology of China, Hefei 230026, People's Republic of China}

\begin{abstract}

We find that the density dependence of the glass transition temperature of Lennard-Jones (LJ) and Weeks-Chandler-Andersen (WCA) systems can be predicted from properties of the zero-temperature ($T=0$) glasses.  Below a crossover density $\rho_s$, LJ and WCA glasses show different structures, leading to different vibrational properties and consequently making LJ glasses more stable with higher glass transition temperatures than WCA ones.  Above $\rho_s$, structural and vibrational quantities of the $T=0$ glasses show scaling collapse.  From scaling relations and dimensional analysis, we predict a density scaling of the glass transition temperature, in excellent agreement with simulation results.  We also propose an empirical expression of the glass transition temperature using structural and vibrational properties of the $T=0$ glasses, which works well over a wide range of densities.

\end{abstract}


\maketitle

According to the theory of liquids, repulsive particle interaction determines the structure of dense liquids, while attraction acts as a perturbation \cite{hansen,wca}.  The perturbative role of attraction has been questioned by recent simulations and experiments through the fact that a glass former with attraction can exhibit slower dynamics and consequently have a higher glass transition temperature than that with purely repulsive interaction even though their structures are similar \cite{berthier1, zhang,pedersen}.  Recent theoretical approaches such as the mode coupling theory have failed to explain such nonperturbative effects of attraction from the liquid side above the glass transition temperature \cite{berthier1}.

Alternatively, some dynamical behaviors of supercooled liquids \cite{debenedetti,berthier2,ediger} can be viewed from the glass side.  For instance, recent studies have shown that the heterogeneous dynamics of supercooled liquids and the glass transition temperature to a great extent reflect the quasi-localized nature of low-frequency normal modes of vibration of the zero-temperature ($T=0$) glasses \cite{widmer-cooper,shintani,wang,manning,chen,ediger,procaccia}.  Can we also gain any insight from studying properties of the $T=0$ glasses to understand the strong dynamical differences between glass formers with attractive and purely repulsive interactions?

We achieve the positive answer to the above question by studying both Lennard-Jones (LJ) and Weeks-Chandler-Andersen (WCA) \cite{wca} systems over a wide range of densities.  We demonstrate that LJ systems with attraction can have higher glass transition temperatures than WCA ones with pure repulsion because LJ glasses have either higher characteristic frequency or weaker quasi-localization of low-frequency modes and are thus more stable \cite{wang,xu1}.  The $T=0$ LJ and WCA glasses have indistinguishable vibrational properties only when their structures are identical.  In order for LJ and WCA systems to have the same glass transition temperature, the $T=0$ glasses must have identical structures, which happens when the density $\rho$ is above a crossover value $\rho_s$.  At $\rho<\rho_s$ where LJ and WCA liquids can have similar structures but different dynamical behaviors, the structures of their $T=0$ glasses are still distinct.

When $\rho>\rho_s$, pair distribution functions of the $T=0$ glasses measured at different densities show scaling collapse.  Correspondingly, vibrational quantities such as the density of vibrational states and mode participation ratio spectrum also show scaling collapse.  From scaling relations and dimensional analysis, we predict a density scaling of the glass transition temperature, which agrees surprisingly well with simulation results.  More interestingly, the dimensional analysis implies an empirical expression of the glass transition temperature in terms of the structural and vibrational information of the $T=0$ glasses.  This expression fits the glass transition temperatures well, even in the density regime where density scaling breaks down.

Our systems are three-dimensional cubic boxes with side length $L$ consisting of $N=1000$ spheres with periodic boundary conditions in all directions.  A mixture of $800~A$ and $200~B$ spheres with equal mass $m$ is employed.  We have verified that there is no significant system size effects.  The potential between particles $i$ and $j$ is \cite{kobandersen}
\begin{equation}
V(r_{ij}) = 4\epsilon_{ij}\left[\left(\frac{\sigma_{ij}}{r_{ij}}\right)^{12} - \left(\frac{\sigma_{ij}}{r_{ij}}\right)^{6}\right] + f(r_{ij}),
\end{equation}
when their separation $r_{ij}$ is smaller than the potential cutoff $r_{ij}^c=\gamma\sigma_{ij}$, and zero otherwise.  Here, $\epsilon_{ij}$ and $\sigma_{ij}$ depend on the type of interacting particles: $\epsilon_{AB}=1.5\epsilon_{AA}$, $\epsilon_{BB}=0.5\epsilon_{AA}$, $\sigma_{AB}=0.8\sigma_{AA}$, and $\sigma_{BB}=0.88\sigma_{AA}$.  $f(r_{ij})$ guarantees that $V(r_{ij}^c)=\frac{{\rm d}V(r_{ij})}{{\rm d}r_{ij}}|_{r_{ij}^c}=0$.  $\gamma$ tunes the range of interaction.  Here we compare systems with $\gamma=2^{1/6}$ (WCA) and $2.5$ (LJ).  We set the length, mass, and energy in units of $\sigma_{AA}$, $m$, and $\epsilon_{AA}$, respectively.  The frequency is thus in units of $\sqrt{\epsilon_{AA}/ m\sigma_{AA}^{2}}$.  The temperature is in units of $\epsilon_{AA} / k_B$ with $k_B$ the Boltzmann constant.

We generate the $T=0$ glasses by quickly quenching equilibrated high-temperature states to local potential energy minima using FIRE minimization algorithm \cite{fire}.  It is verified in Fig.~S5 of the Supplemental Material (SM) that our results do not depend on quench rate.  Normal modes of vibration are obtained by diagonalizing the Hessian matrix using ARPACK \cite{arpack}.  The glass structure is characterized by the pair distribution function of $A$ particles: $g_{_{AA}}(r)=\frac{L^3}{N_A^2}\left< \sum_i\sum_{j\ne i}\delta(r-r_{ij})\right>$, where $N_A$ is the number of $A$ particles and the sums are over all $A$ particles.  Vibrational properties are characterized by the density of vibrational states $D(\omega)=\frac{1}{3N}\left< \sum_l \delta(\omega-\omega_l) \right>$ and participation ratio $p(\omega)=\left<\frac{\sum_l p_l\delta(\omega-\omega_l)}{\sum_l \delta(\omega-\omega_l)} \right>$, where the sums are over all modes, and $p_l=\frac{\left(\sum_{i=1}^N |\vec{e}_{l,i}|^2\right)^2}{N\sum_{i=1}^N |\vec{e}_{l,i}|^4}$ is the participation ratio of mode $l$ with $\vec{e}_{l,i}$ the polarization vector of particle $i$.  Here, $\left< .\right>$ denotes the average over $1000$ independent configurations.

For liquids, we apply molecular dynamics simulations in canonical ensemble and use Gear predictor-corrector algorithm to integrate the equations of motion \cite{allen}.  We measure the intermediate scattering function of $A$ particles, $F_s(k,t)=\frac{1}{N_A}\left<\sum_{j} {\rm exp}({\rm i}\vec{k}\cdot[\vec{r}_{j}(t)-\vec{r}_{j}(0)])\right>$, where the sum is over all $A$ particles, $\vec{r}_j(t)$ is the location of particle $j$ at time $t$, $\vec{k}$ is chosen in the $x-$direction with the amplitude approximately equal to the value at the first peak of the static structure factor, and $\left< .\right>$ denotes ensemble average.  The relaxation time $\tau$ is determined by $F_s(k,\tau)=e^{-1}F_s(k,0)$.  We estimate the glass transition temperature $T_g$ by fitting the relaxation time using Vogel-Fulcher function, $\tau=\tau_0 {\rm exp}\left( \frac{M}{T-T_g}\right)$, where $\tau_0$ and $M$ are fitting parameters.  Fig.~S6 of the SM shows an example about how $T_g$ is measured.

\begin{figure}
\includegraphics[width=0.45\textwidth]{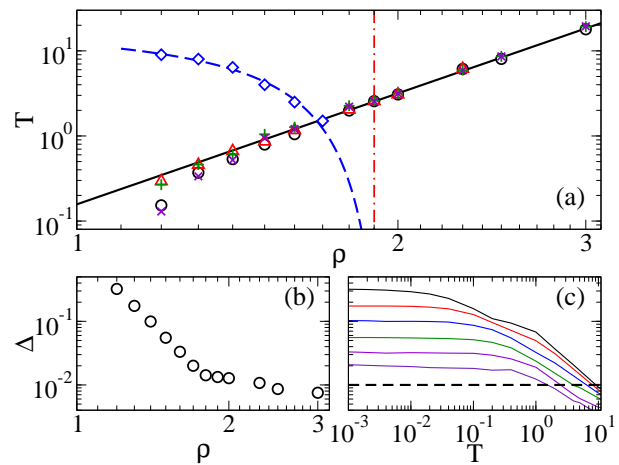}
\caption{\label{fig:fig1} (color online). (a) Density dependence of the glass transition temperatures for WCA (circles) and LJ (triangles) systems, crossover temperature $T_s$ (diamonds), and fits to the glass transition temperature using Eq.~(\ref{Tg_0}) for WCA (crosses) and LJ (pluses) systems.  The solid line shows the density scaling of the glass transition temperature using Eq.~(\ref{Tg}).  The dashed curve is the fit to $T_s$ by $T_s\sim{\rm exp}\left( -\frac{W}{\rho_s^+-\rho}\right)$, where $\rho_s^+\approx 2.0$ and $W\approx 1.1$. The vertical dot-dashed line marks $\rho_s$.  (b) Structural difference $\Delta$ between the $T=0$ LJ and WCA glasses.  (c) Temperature evolution of $\Delta$ between LJ and WCA systems.  From the top to the bottom, the solid curves are for $\rho=1.2$, $1.3$, $1.4$, $1.5$, $1.6$, and $1.7$.  The dashed line marks $\Delta= 0.01$, at which $T_s$ in (a) is estimated.}
\end{figure}

In Fig.~\ref{fig:fig1}(a), we compare the glass transition temperatures of LJ and WCA systems over a wide range of densities.  As already discussed in \cite{berthier1}, $T_g^{LJ}\ge T_g^{WCA}$.  The gap between $T_g^{LJ}$ and $T_g^{WCA}$ shrinks with increasing the density and becomes zero when the density is above a crossover value $\rho_s= 1.9\pm0.1$.

It has been shown that at fixed density structural differences between LJ and WCA systems decrease with increasing the temperature \cite{berthier1}.  We thus calculate $\Delta=\int_0^{\infty} \left| g_{_{AA}}^{WCA}(r) - g_{_{AA}}^{LJ}(r) \right| {\rm d}r$ \cite{prx} to quantify this difference.  In Fig.~\ref{fig:fig1}(b), we first show $\Delta$ at $T=0$.  Interestingly, with increasing $\rho$, $\Delta$ decays to a plateau ($\approx 0.01$, probably due to noises) right at $\rho_s$.  This indicates that $\rho_s$ is the crossover density above which the $T=0$ LJ and WCA glasses have identical structures.  Figure~\ref{fig:fig1}(c) shows the temperature evolution of $\Delta$.  As expected, $\Delta$ decreases with increasing the temperature.  We estimate a temperature $T_s$ at which $\Delta\approx 0.01$ and define it as the crossover temperature above which LJ and WCA systems have identical structures \cite{note}.  As shown in Fig.~\ref{fig:fig1}(a), $T_s$ decreases with increasing the density and can be approximately fitted into $T_s\sim {\rm exp}\left( -\frac{W}{\rho_s^+-\rho}\right)$ with $\rho_s^+\approx 2.0$ and $W\approx 1.1$.  Therefore, $T_s$ decays to zero approximately at $\rho_s$, which implies that above $\rho_s$ LJ and WCA systems have identical structures at all temperatures.  On the contrary, distinct structures of the $T=0$ LJ and WCA glasses at $\rho<\rho_s$ may thus explain why $T_g^{LJ}\ge T_g^{WCA}$.

\begin{figure}
\includegraphics[width=0.47\textwidth]{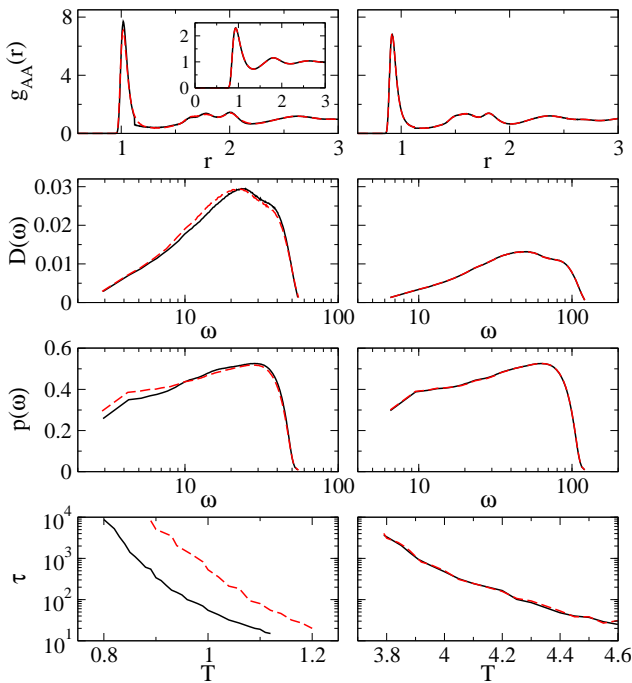}
\caption{\label{fig:fig2} (color online). Pair distribution function $g_{_{AA}}(r)$, density of vibrational states $D(\omega)$, and participation ratio $p(\omega)$ of the $T=0$ glasses, and temperature evolution of the relaxation time $\tau$ of supercooled liquids.  The solid and dashed curves are for WCA and LJ systems, respectively.  The left and right columns are for $\rho=1.4$ and $1.9$, respectively. The inset to the top left panel compares $g_{_{AA}}(r)$ of WCA and LJ liquids at $T=10$.}
\end{figure}

Figure~\ref{fig:fig2} compares the pair distribution function $g_{_{AA}}(r)$, density of vibrational states $D(\omega)$, and participation ratio $p(\omega)$ at $T=0$, and relaxation time $\tau$ at $T>T_g$ between LJ and WCA systems on both sides of $\rho_s$.  The left column is for $\rho=1.4<\rho_s$.  Although $g_{_{AA}}(r)$ of LJ and WCA liquids look identical (see the inset to the top left panel), the $T=0$ structures are apparently different.  The structure plays a key role in determining vibrational properties of glasses.  Since $g_{_{AA}}(r)$ are different, both $D(\omega)$ and $p(\omega)$ of LJ glasses are different from WCA glasses.  In contrast, the right column indicates that $g_{_{AA}}(r)$ of the $T=0$ LJ and WCA glasses at $\rho=1.9\approx \rho_s$ are identical, so do $D(\omega)$ and $p(\omega)$.

In glasses, there are excess low-frequency modes beyond Debye's prediction, which form the boson peak \cite{bosonpeak}.  The boson peak frequency $\omega_{_{BP}}$ at which $D(\omega)/\omega^2$ exhibits a peak has been shown to closely correlate to the stability of glasses \cite{singh}: a glass with lower $\omega_{_{BP}}$ would be less stable.  Modes near $\omega_{_{BP}}$ are usually quasi-localized with small participation ratio \cite{xu1,schober}, {\it i.e.}~a small fraction of less constrained particles vibrate more strongly than the others in the mode.  It has been shown that quasi-localization of low-frequency modes is also strongly responsible to the stability of glasses \cite{xu1}: the energy barrier is lower along the direction of a more localized low-frequency mode.  Therefore, we believe that boson peak frequency and low-frequency quasi-localization interplay to determine the stability of glasses.

\begin{figure}
\includegraphics[width=0.45\textwidth]{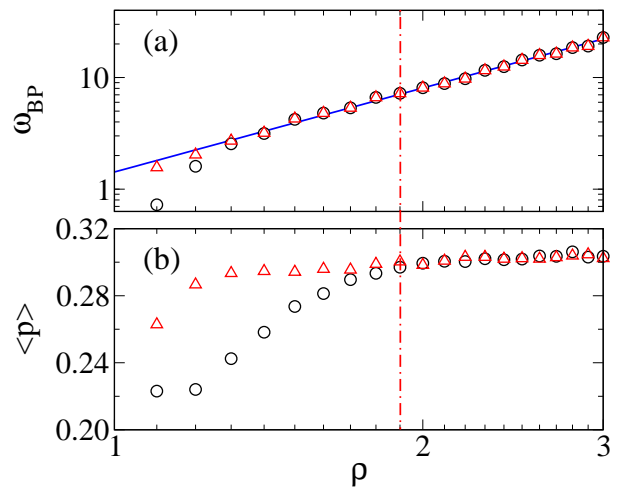}
\caption{\label{fig:fig3} (color online). (a) Boson peak frequency $\omega_{_{BP}}$ and (b) average participation ratio of the lowest $20$ modes $\left< p\right>$ of the $T=0$ WCA (circles) and LJ (triangles) glasses.  The solid line in (a) is the fit to the boson peak frequency using Eq.~(\ref{boson_peak}).  The vertical dot-dashed line marks $\rho_s$.}
\end{figure}

In Fig.~\ref{fig:fig3}, we compare the boson peak frequency $\omega_{_{BP}}$ and average participation ratio over the lowest $20$ modes $\left<p\right>$ for LJ and WCA glasses.  At low densities, {\it e.g.}~$\rho=1.2$ as commonly studied, LJ glasses have larger $\omega_{_{BP}}$ and $\left< p\right>$ than WCA glasses, so it is expected that $T_g^{LJ}>T_g^{WCA}$.  When $\rho_s>\rho>1.4$, $\omega_{_{BP}}$ looks almost identical for both glasses.  However, LJ glasses still have larger $\left< p\right>$ than WCA glasses.  In this density regime, low-frequency quasi-localization plays the key role in determining the difference in stability between LJ and WCA systems.  As illustrated in the bottom left panel of Fig.~\ref{fig:fig2}, supercooled LJ liquids relax more slowly than WCA ones and $T_g^{LJ}>T_g^{WCA}$.  When $\rho>\rho_s$, Fig.~\ref{fig:fig3} indicates that the differences in $\omega_{_{BP}}$ and $\left< p\right>$ between both glasses vanish.  Consequently, $T_g^{LJ}=T_g^{WCA}$ and supercooled LJ and WCA liquids exhibit the same dynamics, as shown in the bottom right panel of Fig.~\ref{fig:fig2}.

\begin{figure}
\includegraphics[width=0.47\textwidth]{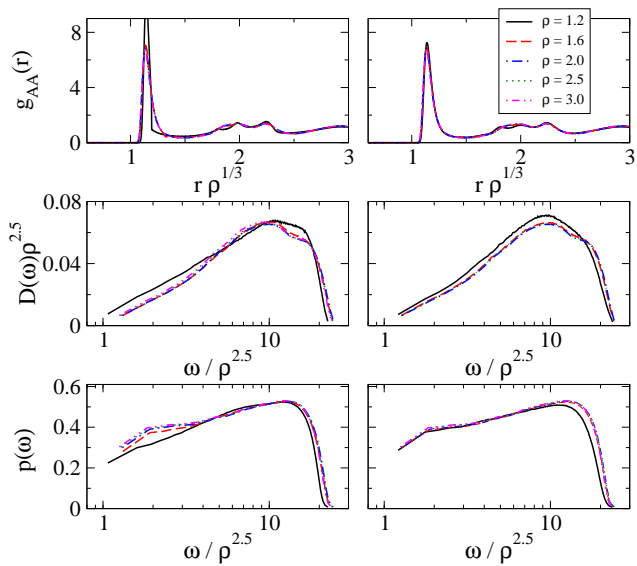}
\caption{\label{fig:fig4} (color online) Scaling collapse of the pair distribution function $g_{_{AA}}(r)$, density of vibrational states $D(\omega)$, and participation ratio $p(\omega)$ for the $T=0$ WCA (left column) and LJ (right column) glasses.}
\end{figure}

The existence of $\rho_s$ clarifies the dynamical differences between LJ and WCA systems.  Moreover, $\rho_s$ is also the crossover density above which $g_{_{AA}}(r)$ of the $T=0$ LJ and WCA glasses show scaling collapse.  As shown in Fig.~\ref{fig:fig4}, $g_{_{AA}}(r)$ measured at $\rho>\rho_s$ collapse onto the same master curve when plotted against $r\rho^{1/3}$.  Similar scaling collapse has also been observed recently for LJ liquids \cite{gnan}.  The scaling collapse implies a characteristic length scale
\begin{equation}
l^*\sim \rho^{-1/3}. \label{length}
\end{equation}

As discussed above, the structure determines vibrational properties of the $T=0$ glasses.  Consequently, both $D(\omega)$ and $p(\omega)$ may show scaling collapse as well.  Figure~\ref{fig:fig3}(a) indicates that at $\rho>1.4$
\begin{equation}
\omega_{_{BP}} \sim \rho^{2.5}, \label{boson_peak}
\end{equation}
which should be the correct scaling relation of the frequency at $\rho>\rho_s$.  In Fig.~\ref{fig:fig4}, we plot $D(\omega)\rho^{2.5}$ and $p(\omega)$ against $\omega/\rho^{2.5}$, and indeed collapse all curves at $\rho>\rho_s$.

The scaling collapse at $\rho>\rho_s$ may contain interesting implications to the potential energy landscape.  In harmonic approximation, frequencies of vibrational modes measure local curvatures of basins of attraction of the $T=0$ glasses.  At $\rho>\rho_s$, potential energy landscapes for LJ and WCA systems may look similar especially near local minima with the average curvatures scaled with $\rho^5$.

From dimensional analysis, we obtain
\begin{equation}
\omega_{_{BP}}\sim \left[\frac{E^*}{m(l^*)^2}\right]^{1/2}, \label{dimension}
\end{equation}
where $E^*$ is a characteristic energy.  At $\rho>\rho_s$, because $\left< p\right>$ is constant in density, $\omega_{_{BP}}$ is the key to determine the stability of glasses, so it is plausible to assume that $E^*\sim T_g$.  Then, the combination of Eqs.~(\ref{length})-(\ref{dimension}) leads to
\begin{equation}
T_g\sim E^*\sim (\omega_{_{BP}}l^*)^2\sim \rho^{13/3}. \label{Tg}
\end{equation}
Interestingly, Eq.~(\ref{Tg}) is exactly the density scaling of the glass transition temperature at $\rho>\rho_s$, as illustrated by the excellent agreement between the the solid line [plot of Eq.~(\ref{Tg})] and simulation results of $T_g$ in Fig.~\ref{fig:fig1}.  For LJ systems, this scaling works even below $\rho_s$, because Eq.~(\ref{boson_peak}) still holds and $\left< p\right>$ still remains constant, as shown in Fig.~\ref{fig:fig3}.  In contrast, Fig.~\ref{fig:fig3} indicates that changes of $\omega_{_{BP}}$ and $\left< p\right>$ in density for WCA glasses at $\rho<\rho_s$ are more pronounced, leading to a faster deviation of the glass transition temperature from Eq.~(\ref{Tg}).

Now let us move one step further.  Seen from the top panels of Fig.~\ref{fig:fig4} and Fig.~S9(b) of the SM, the locations of the first peak and minimum of $g_{_{AA}}(r)$ still follow the scaling described by Eq.~(\ref{length}) when $\rho<\rho_s$, so Eq.~(\ref{length}) is the characteristic length scale for all densities.  Combining Eqs.~(\ref{length}) and (\ref{Tg}) and taking into account the effects of $\left< p\right>$, we propose an empirical expression of the glass transition temperature purely based on structural and vibrational properties of the $T=0$ glasses:
\begin{equation}
T_g\sim \rho^{-2/3}\omega_{BP}^2\left<p\right>. \label{Tg_0}
\end{equation}
As shown in Fig.~\ref{fig:fig1}(a), Eq.~(\ref{Tg_0}) fits all the glass transition temperatures nicely, even at low densities (especially for WCA systems) where density scaling stops working \cite{berthier1}.

Equation~(\ref{Tg_0}) condenses the major findings of our work.  It explains why supercooled LJ and WCA liquids exhibit distinct dynamics from the $T=0$ glass perspective: vibrational properties including the boson peak frequency and low-frequency quasi-localization are distinct, which intrinsically originate from structural differences.  Equation~(\ref{Tg_0}) thus makes a direct quantitative connection between the glass transition and $T=0$ glasses.  The derivation of Eq.~(\ref{Tg_0}) does not require any information of particle interaction, so Eq.~(\ref{Tg_0}) may be generalized to other systems not showing any density scaling at all, which will be discussed elsewhere \cite{wang1}.

Although the crossover density $\rho_s$ is too high to be realized in experiments at present \cite{berthier1}, the physical pictures that it brings about deserve more discussions.  Recent studies have made an interesting point that potential between particles $i$ and $j$ of LJ liquids can be approximated into an inverse power law $V(r_{ij})=4\epsilon_{ij} \left(\sigma_{ij}/r_{ij}\right)^{3\Gamma}+c$ with $c$ a constant varying with $\rho$ \cite{pedersen,gnan,scaling}.  Based on this assumption, $T_g\sim \rho^{\Gamma}$ and relaxation times of LJ liquids can collapse in terms of $\rho^{\Gamma}/T$, which work well at $1.1<\rho<1.4$ with $\Gamma\approx 5.16$ \cite{pedersen,scaling}.  However, as shown in Figs.~S7 and S8 of the SM and discussed in \cite{berthier1,njp}, $\Gamma$ in fact varies with $\rho$, so $\Gamma\approx 5.16$ is an approximation, only working at $1.1<\rho<1.4$.  When $\rho>1.4$, our study indicates that $13/3$ is the correct density scaling exponent, which arises from dimensional analysis and scaling relations of the $T=0$ glasses and cannot be predicted by the inverse power-law potential approximation.  Moreover, Fig.~S7 of the SM shows that the $T=0$ LJ and WCA glasses have different values of $\Gamma$, so the inverse power-law potential approximation cannot be simply applied to predict the existence of $\rho_s$.  More details can be found in the SM.

As suggested by a recent study \cite{prx}, the first coordination shell of Lennard-Jonesian liquids is enough to give correct physics.  One may expect that the first coordination shell is inside the repulsive part of the interaction when $\rho>\rho_s$.  We find that it is not exactly the case.  At $\rho_s$, the first minima of $g_{_{AA}}(r)$ of both LJ and WCA glasses are still beyond the potential minimum at $r=2^{1/6}$ [see Fig.~S9 of the SM].  At densities higher than $2.3$, it is true that the first minimum of $g_{_{AA}}(r)$ lies within the repulsive part of the interaction.  In that regime, behaviors of LJ systems are determined by the repulsive part of the potential, and to distinguish attractive and repulsive inter-particle potentials is not meaningful any more \cite{attractiveforce}.

We are grateful to Ludovic Berthier, Andrea J. Liu, Peng Tan, Gilles Tarjus, and Lei Xu for helpful discussions.  This work was supported by National Natural Science Foundation of China No. 91027001 and 11074228, National Basic Research Program of China (973 Program) No. 2012CB821500, CAS 100-Talent Program No. 2030020004, and Fundamental Research Funds for the Central Universities No. 2340000034.

\section{\emph{\textbf{Supplemental Material}}}

\makeatletter
\def\fnum@figure#1{FIG.~S\thefigure$:$~}
\makeatother

\section{Some simulation information}

\begin{figure}[ht]
\center
\includegraphics[width=0.45\textwidth]{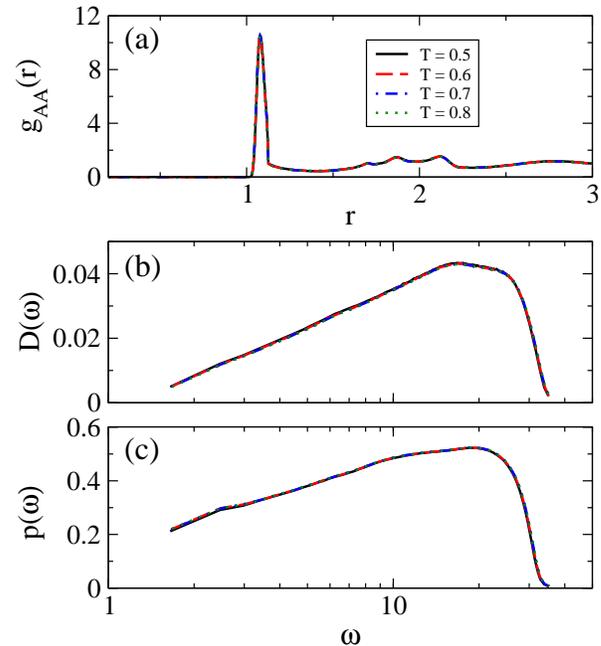}
\caption{\label{sfig:fig1} (a) Pair distribution function of $A$ particles $g_{_{AA}}(r)$, (b) density of vibrational states $D(\omega)$, and (c) participation ratio $p(\omega)$ of the $T=0$ WCA glasses generated from equilibrated states at different initial temperatures.  The density $\rho$ is $1.2$.  The legend shows the values of the initial temperatures.}
\end{figure}

\begin{figure}[ht]
\center
\includegraphics[width=0.48\textwidth]{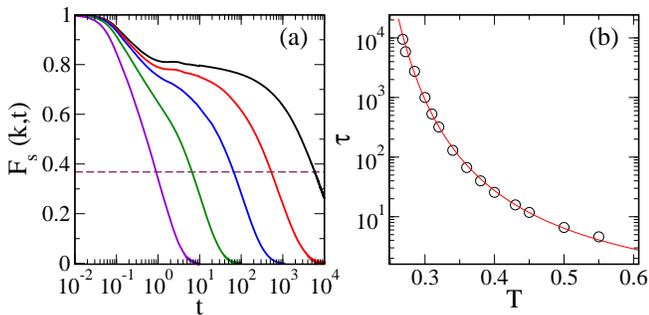}
\caption{\label{sfig:fig2} (a) Intermediate scattering function $F_{s}(k,t)$ for A particles of WCA liquids with $k \sigma_{_{AA}}\simeq 7.25$.  The density $\rho$ is $1.2$. From the left to the right, the temperatures of the liquids are $1.0$, $0.5$, $0.36$, $0.31$ and $0.272$, respectively. The horizontal dashed line marks $F_s = e^{-1}$ at which the relaxation time $\tau$ is determined. (b) Temperature dependence of the relaxation time.  The solid curve is the fit to $\tau$ with Vogel-Fulcher function $\tau=\tau_0 {\rm exp}\left( \frac{M}{T-T_g}\right)$.  For the case shown here, $\tau_0 = 0.19$, $M=1.2$, and $T_{g} = 0.158$.}
\end{figure}

As described in the paper, we generate the $T=0$ glasses by quickly quenching equilibrated high temperature states to local potential energy minima using FIRE minimization method.  The same approach has been widely applied in the study of disordered solids.  In order to check if structural and vibrational properties of the $T=0$ glasses depend on initial condition or quench rate, we prepare equilibrated states at different temperatures above the glass transition temperature and quench them to local potential energy minima.  As shown in Fig.~S\ref{sfig:fig1}, the pair distribution function $g_{_{AA}}(r)$, density of vibrational states $D(\omega)$, and participation ratio spectrum $p(\omega)$ of the $T=0$ glasses do not show apparent initial condition dependence.

Figure~S\ref{sfig:fig2} illustrates an example about how the glass transition temperature is measured.  Figure~S\ref{sfig:fig2}(a) shows the intermediate scattering function $F_s(k,t)$ of WCA liquids measured at different temperatures, from which the relaxation time $\tau$ is obtained.  As shown in Fig.~S\ref{sfig:fig2}(b), the relaxation time increases rapidly with decreasing the temperature, which can be well fitted with the Vogel-Fulcher function as described in the paper.  The glass transition temperature is defined as the extrapolated temperature at which the relaxation time diverges.

\section{Inverse power law potential approximation}

It has been shown that, for strong correlating Lennard-Jones (LJ) liquids with the correlation coefficient between the virial $W$ and potential energy $U=\sum_{ij}V_{ij}$ greater than 0.9, where the sum is over all pairs of interacting particles, fluctuations of the virial and potential energy, $\Delta W$ and $\Delta U$, are linearly correlated: $\Delta W=\Gamma \Delta U$.  Therefore, the pair potential can be well approximated into an inverse power law (IPL) \cite{Sstrongcorrelating1,Sstrongcorrelating2,Spedersen}
\begin{equation}
V_{ij}\approx 4\epsilon_{ij}\left( \frac{\sigma_{ij}}{r_{ij}}\right)^{3\Gamma}+c,
\end{equation}
where $\epsilon_{ij}$, $\sigma_{ij}$, and $r_{ij}$ are defined in the paper, and $c$ is a constant.   As a consequence, dynamics such as the relaxation time $\tau$ of such strong correlating liquids shows density scaling: $\tau (\rho,T) = \rho^{-1/3}T^{-1/2}f\left(\rho^{\gamma}/T\right)$ with $\gamma \approx \Gamma$, which is strictly valid in a pure IPL potential system.  Using this IPL potential approximation, a density scaling with $\gamma\approx 5$ has been proposed, which works well in the vicinity of $\rho=1.2$ \cite{Spedersen,Sscaling1,Sscaling2,Sscaling3,Sscaling4,Sberthier}.

\begin{figure}[ht]
\includegraphics[width=0.48\textwidth]{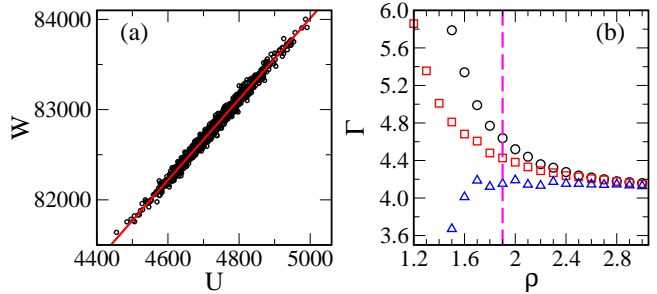}
\center
\renewcommand{\figurename}{FIGS.}
\caption{\label{sfig:fig3} (a) Correlation between the virial $W$ and potential energy $U$ of 1000 $T=0$ LJ glass states at $\rho = 1.8$.  The solid line is the linear fit with a slope of $\Gamma = 4.48$.  (b) Density dependence of $\Gamma$ for the $T=0$ WCA (circles) and LJ (squares) glasses.  The triangles are measured from the very repulsive part of the interaction for the LJ glasses.  The vertical dashed line marks $\rho_{s}=1.9$.    }
\end{figure}

The success of the IPL potential approximation in the vicinity of $\rho=1.2$ for LJ systems may make people believe that behaviors of LJ systems are indeed determined by the approximated IPL potential and $\gamma=5$ is the correct density scaling exponent at even higher densities.  However, there are some doubts that make the above conclusions questionable.  Firstly, in the infinite density limit, the exponent of the very repulsive part of the interaction in the first several coordination shells should be strictly $12$ with $\Gamma=4$.  Apparently, $\gamma=4$ would not collapse relaxation times in the vicinity of $\rho=1.2$.  This simple thinking at least implies that $\Gamma$ is not constant in density, while it varies from around $5$ near $\rho=1.2$ to $4$ in the large density limit.  In fact, the density variation of $\Gamma$ has been demonstrated by Berthier and Tarjus \cite{Sberthier}.  However, they did not move forward to question the validity of the IPL potential approximation.  Secondly, the interaction of the repulsive counterparts of the LJ systems, Weeks-Chandler-Andersen (WCA) systems, can also be well approximated by the IPL potential.  However, it has been shown that density scaling fails to describe the dynamics of WCA systems \cite{Sscaling2,Sberthier}.

Next we will show that the IPL potential approximation indeed has its limitations to predict correct density scaling over a wide range of densities.  Previously reported exponent $\gamma\approx 5$ is only approximately correct in a narrow density range around $\rho=1.2$.  At higher densities, the correct density scaling exponent is $13/3$ obtained from our dimensional analysis of the $T=0$ glasses discussed in the paper, which cannot be correctly predicted from the IPL potential approximation.

\begin{figure}[ht]
\center
\includegraphics[width=0.43\textwidth]{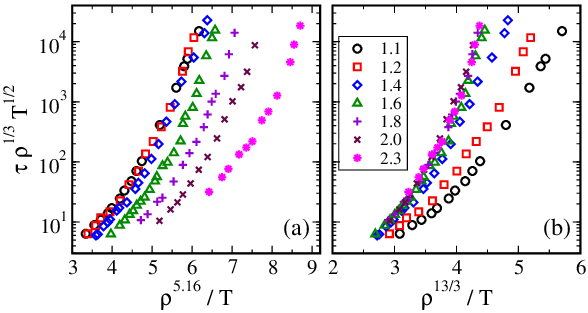}
\caption{\label{sfig:fig4} Scaling of the relaxation time in terms of the temperature and density for LJ liquids.  The scaling exponents used are (a) $5.16$ and (b) $13/3$.  The legend shows values of density. }
\end{figure}

Figure~S\ref{sfig:fig3}(a) shows that fluctuations of the virial and potential energy for the $T=0$ LJ glasses are linearly correlated, from which $\Gamma$ is calculated.  In the paper, we find a crossover density $\rho_s=1.9\pm 0.1$ above which the structural differences between the $T=0$ LJ and WCA glasses vanish, so do the vibrational differences.  It is then easy to assume that interactions of LJ (or its very repulsive part) and WCA glasses can be approximated into the same IPL potential at $\rho_s$.  However, Fig.~S\ref{sfig:fig3}(b) indicates that this is not the case.  Just like what has been discussed above, $\Gamma$ for both LJ and WCA glasses indeed vary continuously with the density.  WCA glasses show a $\Gamma$ larger than LJ glasses.  Only at densities much higher than $\rho_s$, LJ and WCA glasses have the same $\Gamma$.

The density variation of $\Gamma$ thus questions the validity of applying IPL potential approximation and assuming $\gamma\approx\Gamma$ to find the correct density scaling.  This problem has also been realized by B{\o}hling and coworkers who studied densities up to $10$ \cite{Snjp}.  They therefore proposed a new approach instead of the simple IPL potential approximation to describe the dynamics of LJ liquids.

If $\gamma\approx 5$ is indeed a robust density scaling, it should work over a wide range of densities.  However, as shown in Fig.~S\ref{sfig:fig4}(a), the density scaling with $\gamma=5.16$ as proposed by \cite{pedersen} can only collapse relaxation times in a narrow density range $1.1<\rho<1.4$.  It stops working otherwise.  In contrast, Fig.~S\ref{sfig:fig4}(b) shows that the density scaling with $\gamma=13/3$ collapse the data at $\rho>1.4$ nicely.  Therefore, $\gamma=13/3$ obtained from our dimensional analysis is the correct density scaling exponent covering a wide range of densities.  Since $\Gamma$ is not constant in density, $\gamma=13/3$ is not a direct result of the IPL potential approximation.

\section{First Coordination Shell}

\begin{figure}[ht]
\center
\includegraphics[width=0.48\textwidth]{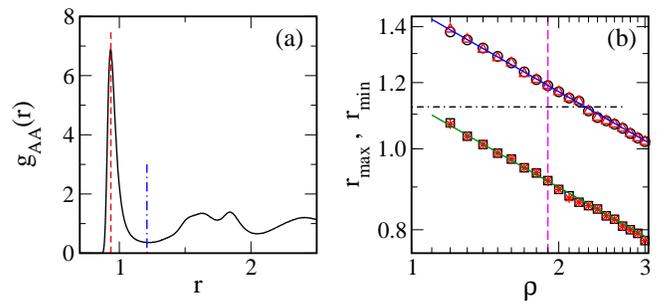}
\caption{\label{sfig:fig5} (a) Illustration of the definition of $r_{max}$ and $r_{min}$ from the pair distribution function $g_{_{AA}}(r)$.  The dashed and dot-dashed lines mark $r_{max}$ and $r_{min}$ at the first peak and minimum of $g_{_{AA}}(r)$.  (b) Density dependence of $r_{max}$ (LJ, stars; WCA, squares) and $r_{min}$ (LJ, circles; WCA, triangles).  The horizontal dot-dashed line marks $r=2^{1/6}$.  The vertical dashed line marks $\rho_s$.  The solid lines have a slope of $-1/3$.}
\end{figure}

For LJ liquids, it has been claimed in \cite{Sprx} that ``in order to get accurate simulation results it is enough to take into account merely the interactions within the FCS if and only if
the liquid is strongly correlating". Here, the first coordination  shell (FCS)  is defined as  the intermolecular interaction range corresponding to the first minimum of
the pair distribution function.  Since above $\rho_{s}$ the structural and vibrational differences between the $T=0$ LJ and WCA glasses vanish, is it possible that $\rho_s$ is also the crossover density above which the FCS for LJ glasses is within the range of repulsion, so that behaviors of the systems are completely determined by repulsion?

We measure the location of the first minimum of $g_{_{AA}}(r)$ of the $T=0$ glasses and denote it as $r_{min}$, as illustrated in Fig.~S\ref{sfig:fig5}(a).  Meanwhile, we locate the first peak of $g_{_{AA}}(r)$ and denote it as $r_{max}$.  As shown in Fig.~S\ref{sfig:fig5}(b), FCS is within the range of repulsion ($r_{min}<2^{1/6}$) when $\rho>2.3$, which is higher than $\rho_s$.  Therefore, there is no necessary connection between $\rho_s$ and the condition that FCS is within the range of repulsion.

Figure~S\ref{sfig:fig5}(b) reveals another interesting phenomenon: both $r_{max}$ and $r_{min}$ are scaled well with $\rho^{-1/3}$ over the whole range of densities studied.  In the paper, we obtain the same density scaling of the characteristic length from scaling collapse of $g_{_{AA}}(r)$ at $\rho>\rho_s$ [see Eq.~(2) of the paper].  Here, Fig.~S\ref{sfig:fig5} demonstrates that this characteristic length scale is present even below $\rho_s$.  This structural scaling becomes one of the three elements for us to propose the empirical expression of the glass transition temperature [Eq.~(6) of the paper].


\begin{thebibliography}{50}
\bibitem{note} Due to noises and accumulated numerical errors, it is impossible and meaningless as well to locate exactly at what temperature $\Delta=0$.  We thus use the plateau value of $\Delta$ at $T=0$ to define $T_s$.  It is already a rather strict criterion.  Below $T_s$ with higher values of $\Delta$, $g_{_{AA}}^{WCA}(r)$ and $g_{_{AA}}^{LJ}(r)$ already look very similar by eye.
\end{thebibliography}
\end{document}